\documentclass[a4paper]{article}

\usepackage{amsmath, amsfonts, hyperref, subcaption, cite, graphicx, epstopdf, geometry}
\usepackage[noblocks]{authblk}
\usepackage[T1]{fontenc}
\usepackage[utf8]{inputenc}

\hypersetup{colorlinks=true,citecolor=cyan, urlcolor=blue,breaklinks}

\author[1,2]{J. L. Alonso}
\affil[1]{Departamento de F\'isica Te\'orica, Universidad de Zaragoza, Pedro Cerbuna 12, ES 50009 Zaragoza, Spain}
\affil[2]{Instituto de Biocomputaci\'on y F\'isica de Sistemas Complejos (BIFI), Universidad de Zaragoza, Mariano Esquillor s/n, Edificio I+D, ES 50018 Zaragoza, Spain}

\author[1,2]{P. Bruscolini}

\author[3]{A. Castro}
\affil[3]{BIFI-Fundaci\'on ARAID, Universidad de Zaragoza, Edificio I+D-Campus R\'io Ebro, Mariano Esquillor s/n, ES 50018 Zaragoza (SPAIN)} 

\author[1,2]{J. Clemente-Gallardo}

\author[4]{J. C. Cuch\'i}
\affil[4]{Departament d'Enginyeria Agroforestal, ETSEA-Universitat de Lleida, Av. Alcalde Rovira Roure 191, ES 25198 Lleida, Spain}

\author[1,2,5]{J. A. Jover-Galtier\thanks{jorge.jover@bifi.es}}
\affil[5]{Centro Universitario de la Defensa de Zaragoza, Academia General Militar, crta. de Huesca s/n, ES 50090 Zaragoza, Spain}

\title{Ehrenfest statistical dynamics in chemistry: study of decoherence effects}

\begin{document}

\begin{center}
{\Large \bf Ehrenfest statistical dynamics in chemistry:\\[0.1cm]study of decoherence effects}\\[1cm]

{\large J. L. Alonso$^{1,2}$, P. Bruscolini$^{1,2}$, A. Castro$^3$, J. Clemente-Gallardo$^{1,2}$,\\[0.1cm]
J. C. Cuch\'i$^{4}$, J. A. Jover-Galtier$^{*1,2,5}$}\\[0.5cm]

{\small \it $^1$ Departamento de F\'isica Te\'orica, Universidad de Zaragoza,\\
Pedro Cerbuna 12, ES 50009 Zaragoza, Spain.\\
$^2$ Instituto de Biocomputaci\'on y F\'isica de Sistemas Complejos (BIFI), Universidad de Zaragoza,\\
Mariano Esquillor s/n, Edificio I+D, ES 50018 Zaragoza, Spain.\\
$^3$ BIFI-Fundaci\'on ARAID, Universidad de Zaragoza, Edificio I+D-Campus R\'io Ebro,\\
Mariano Esquillor s/n, ES 50018 Zaragoza, SPAIN.\\
$^4$ Departament d'Enginyeria Agroforestal, ETSEA-Universitat de Lleida,\\
Av. Alcalde Rovira Roure 191, ES 25198 Lleida, Spain.\\
$^5$ Centro Universitario de la Defensa de Zaragoza, Academia General Militar,\\
ctra. de Huesca s/n, ES 50090 Zaragoza, Spain.
}
\vspace{0.5cm}

*Corresponding author: \href{mailto:jorgejover@unizar.es}{jorgejover@unizar.es}\\[0.1cm]
Accepted for publication in the Journal of Chemical Theory and Computation.\footnote{DOI: \href{https://doi.org/10.1021/acs.jctc.8b00511}{10.1021/acs.jctc.8b00511}}
\end{center}

\begin{abstract}
In previous works, we introduced a geometric route to define our Ehrenfest Statistical Dynamics (ESD) and we proved that, for a simple toy-model, the resulting ESD does not preserve purity. We now take a step further: we investigate decoherence and pointer \emph{basis} in the ESD model by considering some uncertainty in the degrees of freedom of a simple but realistic molecular model, consisting of two classical cores and one quantum electron. The Ehrenfest model is sometimes discarded as a valid approximation to non-adiabatic coupled quantum-classical dynamics because it does not describe the decoherence in the quantum subsystem. However, any rigorous statistical analysis of the Ehrenfest dynamics, such as the described ESD formalism, proves that decoherence exists. In this article, decoherence in ESD is studied by measuring the change in the quantum subsystem purity and by analysing the appearance of the pointer \emph{basis} to which the system decoheres, which for our example is composed by the eigenstates of the electronic Hamiltonian.
\end{abstract}

\section{Introduction}
\label{S1}

The Schr\"odinger equation for a combined system of electrons and nuclei is generally too complex and involves too many degrees of freedom to be solvable, neither analytically nor by numerical methods. Approximations need to be made, one of the most important and successful being the classical approximation for a subset of the particles. Hybrid quantum-classical dynamical (HQCD) models are therefore necessary and widely used \cite{Alonso2011, Alonso2012a, Bedard2005JCP, Landry2011JCP, Larsen2006, Prezhdo1999JCP, Subotnik2010, Subotnik2011a, Subotnik2011b, Sun1998a, Truhlar2007Book, Tully1990JCP, Tully1998a, Tully1998Book, Yonehara2012, zhu2005non}. In a previous work \cite{Alonso2012a} we discussed how these HQCD models are built. Most approaches can be described in two steps: first, a partial deconstruction of the quantum mechanics (QM) of the total system (electrons and nuclei) which simplifies the model, and then a reconstruction that aims to recover the essential properties of the total Schr\"odinger equation lost in the deconstruction process.

HQCD models in the literature present at least two levels of deconstruction. The first one, called Born-Oppenheimer molecular dynamics (BOMD), is far away from the total Schr\"odinger equation for electrons and nuclei, as electrons are assumed to remain in the ground state for all times. The second one, closer to the total Schr\"odinger equation, is called Ehrenfest Dynamics (ED). In ED the nuclei are still classical (as in BOMD) but the electrons are allowed to populate excited states. A recent review on the topic \cite{Yonehara2012} discusses these two approaches. The deconstruction simplifies the model by forcing the separability of the quantum states of the nuclei (which are later considered as classical) and the electrons \cite{Bornemann1996}, even if this separability cannot be preserved exactly by the evolution of an interacting quantum system. Therefore the deconstruction ignores entangled states of the quantum nuclear and electronic degrees of freedom. One could naively conclude that this implies a unitary evolution for the electronic subsystem, thus being the cause for the preserved purity in the ED evolution. This reasoning is erroneous, as generic ED evolution is not unitary for the electronic subsystem\cite{Alonso2012a}. Nevertheless, as proved in Theorem 1 of the cited work, ED does preserve purity, and no decoherent effects can be found.

The second step in the definition of HQCD models, the reconstruction of the dynamics, is much more complicated. Many different proposals tackle the difficulty of re-incorporating into HQCD models the essential properties of the total Schr\"odinger equation that have been lost. One of these properties is the decoherence phenomenon in the electronic subsystem. In the context of HQCD, decoherence is understood as the fact that the neglected wave functions in the classical limit (i.e. those of the nuclei) rapidly lose overlap in time, leading to the destruction of superpositions in the quantum subsystem, i.e. forcing the electronic wave functions into a mixture of pure states \cite{Bedard2005JCP, Larsen2006}. Such pure states form the so-called \emph{pointer basis} for the quantum sybsystem \cite{Dannenberg2011, Lychkovskiy2009, Schlosshauer2007}.

Different approaches aim to address the reconstruction of quantum dynamics with quite different tools. In J.~C. Tully’s Trajectory Surface Hopping (TSH) algorithms\cite{Tully1990JCP}, for example, the deconstruction goes to BOMD and the reconstruction proceeds by allowing the system to perform certain specially designed stochastic jumps between adiabatic states. These jumps cannot however be well justified from first principles. Another relevant algorithm, widely used in Molecular Dynamics (MD), is the decay of mixing formalism by Truhlar and coworkers \cite{zhu2005non, Truhlar2007Book}. In this method, the deconstruction stops at the ED and the reconstruction is developed by adding terms in the dynamics which introduce decoherence. In this formalism, one considers an ensemble of hybrid quantum-classical systems and computes the dynamics of the quantum subsystem using two components: one arising as the fully coherent solution to the Liouville-von Neumann equation and one, \emph{ad-hoc}, that incorporates electronic decoherence (see expression \eqref{dissipTruhlar} below).

There are other studies and proposals to tackle decoherence effects. Among them, we would like to mention Bittner and Rossky \cite{Bittner1995}, Neria and Nitzan \cite{Neria1993}, Schwartz and coworkers \cite{Schwartz1996} and Subotnik \cite{Subotnik2010, Subotnik2011a}. Particularly in the last two approaches, one can find the idea of considering statistical mixtures of hybrid quantum-classical systems in order to study the problem of decoherence. For example, Subotnik used in his works the formalism of the partial Wigner transform, introduced in the context of MD by Nielsen and coworkers \cite{Kapral1999a, Nielsen2000, Nielsen2001a}, to represent the hybrid quantum-classical system; by adding some extra variables into the picture, it is possible to describe in an efficient way the decoherence effects of some systems.

A completely different route is the one followed by Abedi and coworkers \cite{Abedi2010, Abedi2012, Abedi2013, Alonso2013}, which prescribes an exact factorisation of the total wave function. Then, the classical limit for the nuclear part can be taken, and recently the appearance of decoherence in the resulting dynamics has been analyzed \cite{Min2017}.

In this work, we take a step back from these approaches, and examine how uncertainty in the initial conditions may affect Ehrenfest dynamics. In order to adress this issue, a formulation of dynamics in terms of statistical distributions is necessary. In previous works \cite{Alonso2011, Alonso2012a} we introduced a geometric route to define our Ehrenfest Statistical Dynamics (ESD). We now take a step further by investigating decoherence in ESD. It is important to notice that, due to the interaction between classical cores and quantum electrons in a molecular model, cores act as an environment to the quantum subsystem of the molecule. In this setting, it is thus possible to consider the \emph{decoherence hypothesis} \cite{Dannenberg2011, Lychkovskiy2009, Schlosshauer2007}, by which the environment of a non-isolated quantum system selects a set of orthogonal projectors onto the Hilbert space of the system. In other words, the interaction described in Ehrenfest dynamics causes the existence of a pointer basis for the quantum subsystem.

In our previous work \cite{Alonso2012a} we proved that, for a simple toy-model,the resulting dynamics does not preserve purity. Now, and after carefully defining the meaning of decoherence in the context of HQCD models, we apply ESD to a realistic model: a diatomic, isolated molecular system. We will see that decoherence does occur in ESD and we will compute the changes in purity and the appearance of pointer basis associated to the decoherence phenomenon.

In Section \ref{S2} we summarise the main properties of ED and the corresponding statistical model. We also detail in Section \ref{S2.3} how ESD defines an evolution for the quantum subsystem which does not preserve its purity; we define the corresponding decoherence-time and the pointer basis. Section \ref{sec:exampl-dynam-simul} studies the dynamics of a ionised dimer subject to uncertainty in its initial conditions. By making use of ESD, we prove the appearance of decoherence and we determine the pointer basis for this example, which turns out to be the set of eigenstates of the electronic Hamiltonian. Conclusions and an outlook of the work are presented in Section \ref{secConcl}.

\section{Ehrenfest Dynamics and Statistics}
\label{S2}

\subsection{The case of molecular systems: Ehrenfest model}
\label{S2.1}

In Ehrenfest Dynamics (ED), the particles in a molecule are split in two sets \cite{Horsfield2006}: 
\begin{itemize}
\item $N_Q$ quantum particles, typically the most external (valence) electrons,
\item which are coupled to the $N_C$ cores that will be considered classical.
\end{itemize}

In the following, the coordinates of electrons and cores will be denoted by 3-dimensional vectors $\vec{r}_1, \ldots, \vec{r}_{N_Q}$ and $\vec{R}_1, \ldots, \vec{R}_{N_C}$, respectively. The Hamiltonian operator for this molecular system is given by (atomic units will be used hereafter)
\begin{equation}
H = - \sum_{J=1}^{N_C} \frac{1}{2 M_J} \nabla_J^2 + H_e (R),
\end{equation}
with $R = (\vec{R}_1, \ldots, \vec{R}_{N_C}) \in \mathbb{R}^{3N_C}$, and
where $M_J$ is the mass of the $J$-th core and $\nabla_J^2$ is the Laplacian operator with respect to $\vec{R}_J$. The operator $H_e (R)$, called the electronic Hamiltonian of the molecule, depends parametrically on the core positions:
\begin{equation}
\label{eq:29}
\begin{aligned}
H_e(R) = & - \frac{1}{2} \sum_{j=1}^{N_Q} \nabla_j^2
+ \sum_{J<K} \frac{Z_J Z_K}{|\vec R_J - \vec R_K|} + \sum_{j<k} \frac{1}{|\vec r_j-\vec r_k|}
- \sum_{J,j}\frac{Z_J}{|\vec R_J-\vec r_j|},
\end{aligned}
\end{equation}
where $Z_J$ is the charge of the $J$-th core and $\nabla_j^2$ is the Laplacian operator with respect to electronic coordinates $\vec{r}_j$. 

The first approximation to the solution of the otherwise unassailable Schr\"odinger equation for this Hamiltonian is the separability of electrons and cores:
\begin{equation}
\label{eq:19}
|\Phi\rangle = |\xi\rangle \otimes |\psi\rangle.
\end{equation}
with $|\xi \rangle$ representing the state of the cores and $|\psi\rangle$ the electronic state. This separability assumption ignores the possibility of electron-nuclear entanglement\cite{Horsfield2008}.

From these separable states, the next step is the substitution of the wave function for the cores by single classical trajectories, characterised by positions $R = (\vec{R}_1, \ldots, \vec{R}_{N_C} ) \in \mathbb{R}^{3N_C}$ and momenta $P = (\vec{P}_1, \ldots, \vec{P}_{N_C} ) \in \mathbb{R}^{3N_C}$ variables. The states in the hybrid quantum-classical model are then represented by elements in the following spaces:
\begin{itemize}
\item the phase space of the $N_C$ classical cores corresponds to
\begin{equation}
M_C= \overbrace{\mathbb{R}^{6} \times \cdots \times \mathbb{R}^{6}}^{N_{C}} = \mathbb{R}^{6N_C}.
\end{equation}
\item the quantum Hilbert space $\mathcal{H}$ describing the states of the most external electrons.
\end{itemize}

The original Schr\"odinger equation of the full quantum model is approximated by a set of coupled differential equations when written on the hybrid quantum-classical variables introduced above \cite{Tully1998Book, Tully1998a, Bornemann1996, Jover-Galtier2017, Marx2000, Marx2009}. These are called the Ehrenfest equations. For a system composed of a set of $N_{C}$ classical particles, described by points $\xi = (R,P) \in M_C$, and a quantum subsystem, described by a wavefunction $|\psi\rangle \in \mathcal{H}$, the Ehrenfest equations are:
\begin{equation}
\label{eq:ehrenfest}
\begin{aligned}
& \frac{d}{dt} \vec{R}_J(t) = \frac{\vec{P}_J}{M_J}, \\
& \frac{d}{dt} \vec{P}_J(t) = -\left\langle \psi(t) \left\vert \frac{\partial H_e}{\partial \vec R_J}(R(t)) \right\vert \psi(t) \right\rangle, \\
& i\hbar \frac{d}{dt} |\psi(t)\rangle = H_e(R(t))\vert\psi(t)\rangle.
\end{aligned}
\end{equation}

\subsection{Hybrid quantum-classical Eh\-ren\-fest statistical model} 
\label{S2.2}

\subsubsection{Hybrid mechanical systems}

In previous works\cite{Alonso2011,Alonso2012a}, we introduced a geometric formulation of ED for hybrid quantum-classical models and its extension to Statistical Mechanics. The first step was to adapt the geometrical formulation of Quantum Mechanics \cite{Brody2001, Carinena2007b, ClementeGallardo:2008p614, Heslot1985a, Jover-Galtier2017, Kibble:1979p7279, Meyera1979} for the description of hybrid quantum-classical systems.

The main idea of the construction is to combine the classical and the quantum degrees of freedom in a form similar to the composition of classical systems. This is achieved by providing a description of the quantum subsystem formally equivalent to the description of the classical one. Let us take any basis $\{ |e_j \rangle\}_j$ of the Hilbert space ${\mathcal H}$ and consider the real and imaginary parts of the corresponding set of coordinates:
\begin{equation}
\label{eq:1}
|\psi\rangle = \sum_j z_j |e_j\rangle, \ z_j = \frac{1}{\sqrt{2}} \left( q_j + i p_j \right).
\end{equation}
The real numbers $(q_1, p_1, q_2, p_2, \cdots )$ can be understood as coordinates on a real differentiable manifold\footnote{Infinite-dimensional Hilbert spaces cannot be given a consistent geometric description. In practice, this problem is avoided since the full Hilbert spaces are substituted by truncated discretised spaces.} $M_Q$, whose points are in one-to-one correspondence with the vectors in $\mathcal{H}$. This is a K\"ahler manifold \cite{Brody2001, Carinena2007b, ClementeGallardo:2008p614, Heslot1985a, Jover-Galtier2017, Kibble:1979p7279}, with a Poisson bracket $\{\cdot, \cdot\}_Q$ and a symmetric product of functions $(\cdot,\cdot)_Q$ determined by the properties of the Hermitian product in $\mathcal{H}$. Their coordinate expressions are
\begin{equation}
\label{prodsQ}
\begin{aligned}
\{f, g\}_Q = & \sum_j \left( \frac{\partial f}{\partial q_j} \frac{\partial g}{\partial p_j} - \frac{\partial f}{\partial p_j} \frac{\partial g}{\partial q_j} \right), \quad (f, g)_Q = \sum_j \left( \frac{\partial f}{\partial q_j} \frac{\partial g}{\partial q_j} + \frac{\partial f}{\partial p_j} \frac{\partial g}{\partial p_j} \right).
\end{aligned}
\end{equation}

Schr\"odinger's equation can be written as a system of Hamilton equations with respect to the quantum Poisson bracket $\{\cdot, \cdot\}_Q$ \cite{Kibble:1979p7279, Brody2001, Heslot1985a, Carinena2007b, ClementeGallardo:2008p614, Jover-Galtier2017}. Therefore, a purely quantum system can be considered as a ``classical'' Hamiltonian system with respect to the symplectic structure on the manifold $M_Q$.

This geometric framework allows to combine the quantum and the classical parts of Ehrenfest's equations in a Hamiltonian system defined on the product manifold $M_{C}\times M_{Q}$, where $M_{C}$ is the phase space of the classical degrees of freedom defined above. The composition is completely analogous to the composition of two classical systems: the phase space of the whole system is the Cartesian product of the phase spaces of the subsystems, and a global Poisson bracket (or equivalently a global symplectic structure) is obtained by adding those of the subsystems:
\begin{equation}
\label{eq:27}
\{ \cdot, \cdot\}_{QC}=\{ \cdot, \cdot\}_{C}+\hbar^{-1}\{ \cdot, \cdot\}_{Q},
\end{equation}
where $\{ \cdot, \cdot\}_{C}$ is the canonical classical Poisson bracket on $M_C$.

In a geometrical formalism, the physical observables are represented by functions on the global phase space of the system. From the geometric formalism of Quantum Mechanics, we know that a purely quantum operator, identified as a Hermitian operator $A$ on the Hilbert space of the quantum system, is represented geometrically by the function
\begin{equation}
\label{eq:2}
f_{A}(\psi)= \langle\psi |A |\psi\rangle.
\end{equation}
Hamiltonian dynamics associated to these functions preserve the K\"ahler structure on the manifold $M_Q$ \cite{Kibble:1979p7279, Brody2001, Heslot1985a, Carinena2007b, ClementeGallardo:2008p614, Jover-Galtier2017}, represented by the products of functions in \eqref{prodsQ}. This property should also be true for generic observables on quantum-classical systems. As a consequence, for each $\xi \in M_C$ any observable has to be represented by a Hermitian operator $A (\xi)$ acting on the Hilbert space of the quantum subsystem \cite{Heslot1985a}. Thus, a function $f_{A}(\xi, \psi)$ representing a hybrid observable can always be written as
\begin{equation}
\label{fAHybr}
f_A (\xi, \psi) = \langle\psi |A(\xi) |\psi\rangle.
\end{equation}
When introducing mixed quantum-classical observables, it is relevant to analyse their algebraic structures. In Classical Mechanics, observables are represented by smooth functions on the phase space, which close a Lie algebra with respect to the classical Poisson bracket. For observables on quantum systems, as first shown by Heslot \cite{Heslot1985a}, it is enough to consider functions of the form of \eqref{eq:2}, which also close a Lie algebra, in this case with respect to the quantum Poisson bracket $\{\cdot, \cdot\}_Q$ defined in \eqref{prodsQ}. In the case of mixed quantum-classical systems, the natural extension of quantum functions, represented by \eqref{fAHybr}, no longer close a Lie algebra with respect to the new Poisson bracket \eqref{eq:27}. Indeed, its action over two such functions, which are quadratic on the quantum degrees of freedom, would in general produce a non-quadratic function. Thus, from an algebraic point of view, a Lie algebra with respect to the quantum-classical Poisson bracket is obtained by considering all the smooth functions on $M_{QC}$, as already proposed by some of us \cite{Alonso2011}. Observe that only the subset of functions of the form \eqref{fAHybr} corresponds to physical observables, while all the remaining functions have in principle no physical meaning.

The function $f_H$ associated to the molecular Hamiltonian represents the energy of the hybrid quantum-classical system \cite{Alonso2011, Alonso2012a}. It is given by the expression:
\begin{equation}
\label{eq:4}
\begin{gathered}
f_H (\xi, \psi) = \sum_{J=1}^{N_C} \frac{P_J^2}{2 M_J} + f_{H_e} (\xi, \psi), \quad
f_{H_e} (\xi, \psi) = \langle \psi |H_{e}(R)|\psi\rangle.
\end{gathered}
\end{equation}
The electronic Hamiltonian $H_{e}(R)$ for finite-dimensional systems is obtained in an analogous way to \eqref{eq:29}. By using function $f_H$ and the Poisson bracket \eqref{eq:27}, it is possible to define a Hamiltonian vector field whose integral curves are the solutions to the Ehrenfest equations \cite{Alonso2011}. 

\subsubsection{Ehrenfest statistical systems}

It can be concluded that Ehrenfest equations, written in the appropriate way, are formally analogue to a classical Hamiltonian system. Therefore, from Liouville theorem we can define a volume on $M_C \times M_Q$ which is invariant under the dynamics. This allows us to define a statistical mechanical model where Ehrenfest equations define the dynamics of the microstates. The statistical description is defined by a distribution density function $F_{QC}$ on the manifold $M_C \times M_Q$ satisfying the following normalisation condition:
\begin{equation}
\label{eq:normalisation}
\begin{gathered}
\int_{M_C \times M_Q} d \mu_{QC} (\xi, \psi) F_{QC} (\xi, \psi) = 1, \quad
d \mu_{QC} (\xi, \psi) = d \mu_C(\xi) \otimes d \mu_Q (\psi),
\end{gathered}
\end{equation}
where $d\mu_{QC} (\xi, \psi)$ represents the volume element on $M_C\times M_Q$ defined by the symplectic volumes $d\mu_C (\xi)$ and $d\mu_Q (\psi)$ on the classical and quantum phase spaces, respectively.

From an empirical perspective, the description of molecular systems by means of probability distributions seems natural, since any macroscopic system will have a certain distribution of states, associated to thermal motion or to a simple uncertainty of the exact state of the particles. Notice that, as we have a distribution depending on two types of degrees of freedom, we can consider the corresponding marginal distributions defined as:
\begin{equation}
\label{eqFCFQ}
\begin{aligned}
F_C (\xi) = & \int_{M_Q} d\mu_Q (\psi) F_{QC} (\xi, \psi), \quad \xi \in M_C, \\
F_Q (\psi) = & \int_{M_C} d\mu_C (\xi) F_{QC} (\xi, \psi), \quad \psi \in M_Q.
\end{aligned}
\end{equation}
Marginal distributions $F_{C}$ and $F_{Q}$ encode respectively the corresponding uncertainties of the classical and quantum degrees of freedom. 

By using these probability distribution we can measure the average value of any magnitude, classical, quantum or hybrid. The statistical average of a hybrid observable $A$, represented by a function $f_{A}(\xi, \psi)$, is obtained as
\begin{equation}
\label{eq:3}
\langle A \rangle = \int_{M_C \times M_Q} d\mu_{QC} (\xi, \psi) F_{QC} (\xi, \psi) \frac{f_A (\xi, \psi)} {\langle \psi|\psi \rangle}.
\end{equation}

We can also define a density matrix to represent the quantum subsystem in a more standard way \cite{Alonso2012a} by averaging the projectors on the quantum states by the distribution $F_{QC}$: 
\begin{equation}
\label{eq:5}
\begin{aligned}
\rho = & \int_{M_C \times M_Q} d\mu_{QC} (\xi, \psi) F_{QC} (\xi, \psi) \frac{|\psi\rangle \langle \psi|}{\langle \psi|\psi \rangle}
= \int_{M_Q} d\mu_Q (\psi) F_Q (\psi) \frac{|\psi\rangle \langle \psi|}{\langle \psi|\psi \rangle}.
\end{aligned}
\end{equation}
The last equality shows that the density matrix $\rho$ is the representation of the marginal distribution $F_Q$ as a density operator. It is immediate to verify that the density matrix $\rho$ allows us to write the average value of a pure quantum observable $B$ as
\begin{equation}
\label{eq:5bc}
\langle B \rangle = \mathrm{Tr} (\rho B).
\end{equation}

Analogously, we can consider a $\xi$-dependent operator in order to represent in the same language the total probability distribution $F_{QC}$:
\begin{equation}
\label{eq:21}
\rho_C (\xi) = \int_{M_Q} d\mu_Q (\psi) F_{QC} (\xi, \psi) \frac{|\psi\rangle \langle \psi|}{\langle \psi|\psi \rangle}.
\end{equation}
Observe that this is operator is not a density matrix, as in general ${\rm Tr} \rho_C (\xi) \neq 1$. It is possible to relate \eqref{eq:5} and \eqref{eq:21} as
\begin{equation}
\label{eq:32}
\rho = \int_{M_C} d\mu_C (\xi) \rho_C (\xi).
\end{equation}

\subsubsection{Simple examples}

The expressions obtained above are valid for any generic probability distribution $F_{QC}$. 
A ``pure state'', in which the state of both the classical and quantum subsystem are completely determined 
to be $\xi_0$ and $\psi_0$, corresponds to:
\begin{equation}
\label{FQCsimple}
F_{QC} (\xi, \psi) = \delta (\xi - \xi_0) \delta(\psi - \psi_0),
\end{equation}
The corresponding $\rho_C(\xi)$ operator, defined by \eqref{eq:21}, is
\begin{equation}
\label{eq:24}
\rho_C (\xi) = \delta (\xi - \xi_0) \frac{|\psi_0 \rangle \langle \psi_0|}{\langle \psi_0|\psi_0 \rangle}. 
\end{equation}

To consider some uncertainty on the state of the molecule, we may study a situation such as
\begin{equation}
\label{FQCuncertXi}
F_{QC} (\xi, \psi) = f(\xi) \delta (\psi - \psi_0),
\end{equation}
that describes a molecule with uncertainty in the classical degrees of freedom, described by a probability distribution $f(\xi)$ on $M_C$. This is the case studied in a previous work \cite{Alonso2012a} for a simplified model where the distribution was chosen as
\begin{equation}
\label{eq:26}
f (\xi) = \frac 1N \sum_{j=1}^N \delta (\xi - \xi_j).
\end{equation}
With the definition in \eqref{eq:21}, the $\rho_C(\xi)$ operator associated to \eqref{FQCuncertXi} is
\begin{equation}
\label{eq:25}
\rho_C (\xi) = f(\xi) \frac{|\psi_0\rangle \langle\psi_0|}{\langle \psi_0|\psi_0 \rangle}
\end{equation}

\subsubsection{General systems}

The choice of discrete distributions for the quantum degrees of freedom may seem a very special case, but as we are going to see it covers all possible situations. The reason for that arises from Gleason's theorem \cite{Gleason1957} and the given definition for $\rho_C(\xi)$ in \eqref{eq:21}. Gleason proved that the state of any quantum system can be encoded under the form of a density matrix. Extending this theorem to the case of hybrid quantum-classical systems, for a fixed value of the classical degrees of freedom $\xi_*$, the density $F_{QC}(\xi_{*}, \psi)$ must determine an operator $\rho_C(\xi_{*})$ with the following properties:
\begin{itemize}
\item it is Hermitian: $\rho_C (\xi_*)^{\dagger} = \rho_C (\xi_*)$;
\item it is positive: $\rho_C (\xi_*) > 0$.
\end{itemize}

Normalisation \eqref{eq:normalisation} of the full probability density $F_{QC}$ implies that the family of operators $\rho_C (\xi)$ satisfies
\begin{equation}
\label{eq:53}
\int_{M_C} d\mu (\xi) \mathrm{Tr} \rho_C (\xi) = 1.
\end{equation}

It is immediate to verify that expression \eqref{eq:3} for the average value $\langle A\rangle$ of a hybrid observable $A$ can be written as
\begin{equation}
\label{eq:51}
\langle A \rangle = \int_{M_C} d\mu (\xi) \mathrm{Tr} \left( \rho_C (\xi) A(\xi) \right).
\end{equation}

It is also immediate to understand that there are multiple probability densities which produce the same result for all possible observables on the system. Indeed, consider a generic probability density $F_{QC}$ and its associated operator $\rho_C (\xi)$. It is possible to consider the spectral decomposition of this operator, which allows us to rewrite it as
\begin{equation}
\label{eq:54}
\rho_C(\xi) = \sum_{k=1}^n \lambda_k (\xi) \frac{|\chi_k(\xi) \rangle\langle \chi_k(\xi)|}{\langle \chi_k(\xi)|\chi_k(\xi) \rangle},
\end{equation}
where $\lambda_k(\xi)$ is the $k$-th eigenvalue of the operator and $\chi_k(\xi)$ represents the corresponding eigenvector. We can now define a new probability density as
\begin{equation}
\label{eq:52}
F'_{QC} (\xi, \psi) = \sum_{k=1}^n \lambda_k (\xi) \delta(\psi - \chi_k(\xi)).
\end{equation}
The probability densities $F_{QC}$ and $F'_{QC}$ define the same operator $\rho_C (\xi)$, and therefore they produce the same expectation values for any hybrid observable of the system. In other words, these probability densities are indistinguishable and physically equivalent.

We can conclude that, for hybrid systems with finite dimensional quantum subsystems, it suffices to consider a family of discrete quantum distributions indexed by the classical degrees of freedom as in \eqref{eq:52} to cover all physically meaningful situations. In the analysis of statistical systems, however, it is useful to consider both approaches, as each one has its advantages: the operators $\rho_C (\xi)$ are biunivocally related to the state of hybrid quantum-classical systems, whereas the probability density $F_{QC}$ can be more convenient for some mathematical manipulations, e.g. using Liouville's theorem.

\subsection{Decoherence in Ehrenfest Statistical Dynamics}

\subsubsection{Dynamical evolution}

As seen above, the geometric formalism shows us that ED is of Hamiltonian type \cite{Alonso2011}, which implies that the volume element $d \mu_{QC}$ is preserved by the evolution. As a consequence, in the case of statistical systems, the evolution can be translated onto the space of probability distributions by means of the Liouville equation \cite{Alonso2012a, Balescu1975, Balescu1997}:
\begin{equation}
\label{eq:10}
\frac{\partial}{\partial t} F_{QC} = -\{ f_H, F_{QC} \}_{QC},
\end{equation}
where $f_H$ is given for the Ehrenfest model by \eqref{eq:4} and $\{\cdot, \cdot \}_{QC}$ represents the Poisson bracket on $M_C \times M_Q$ is defined by \eqref{eq:27}. While the Ehrenfest equation for pure systems obviously cannot account for the phenomenon of decoherence, this statistical Ehrenfest equation does allow for decoherence to take place, as we will now show.

The time-dependent density matrix of the system is given by
\begin{equation}
\label{eq:11}
\rho (t) = \int_{M_C \times M_Q} d\mu_{QC} (\xi, \psi) F_{QC}(\xi, \psi; t) \frac{|\psi\rangle \langle \psi|}{\langle \psi|\psi \rangle}.
\end{equation}
Liouville's equation \eqref{eq:10} determines the evolution of the time-dependent probability density $F_{QC}(\xi, \psi; t)$ and from it we can immediately recover the density matrix for any time. In the case of a pure system:
\begin{equation}
F_{QC} (\xi, \psi; 0) = \delta (\xi - \xi_0) \delta (\psi - \psi_0)\,,
\end{equation}
the evolution of the corresponding density matrix is given by:
\begin{equation}
\rho (0) = \frac{|\psi_0\rangle \langle \psi_0|}{\langle \psi_0|\psi_0 \rangle}, \
\dot \rho(t) = -i \hbar^{-1} [H_e (R(t)), \rho(t)],
\end{equation}
which resembles the expression of von Neumann's equation for quantum systems. The dependence on the classical positions $R(t)$, however, causes this evolution to be non-linear, and therefore non-unitary. Nevertheless, the purity is still preserved, as this is simply a rewriting of the equations of the standard Ehrenfest model \eqref{eq:ehrenfest}.

However, more complex distributions will evolve following more complex equations, decoherence will appear, von Neumann's equation will no longer hold, and purity will no longer be preserved. Consider that we add an extra term to our previous pure state, in the form:
\begin{equation}
\label{eq:20}
\begin{aligned}
F_{QC} (\xi, \psi) = & \delta (\xi - \xi_0) \delta (\psi - \psi_0)
+ \tilde F_C (\xi) \delta (\psi - \psi_0),
\end{aligned}
\end{equation}
for a generic function $\tilde F_C(\xi)$, with the only constraint of having zero integral, in order to satisfy the normalisation condition \eqref{eq:normalisation} for $F_{QC} (\xi, \psi)$. The dynamics for the density matrix of the quantum subsystem is
\begin{equation}
\label{eq:22}
\begin{aligned}
\dot \rho(t) = & -i \hbar^{-1} \left[ H_e (\xi (\xi_0, \psi_0,t)),  \frac{|\psi (\xi_0, \psi_0, t) \rangle \langle \psi(\xi_0, \psi_0, t)|}{\langle \psi (\xi_0, \psi_0, t) | \psi(\xi_0, \psi_0, t) \rangle}\right] \\
& - i\hbar^{-1} \int_{M_C} d\mu_{C} (\xi') \tilde F_C (\xi') \left[ H_e (\xi (\xi', \psi_0,t)),  \frac{|\psi (\xi', \psi_0, t) \rangle\langle \psi (\xi', \psi_0, t)|}{\langle \psi (\xi', \psi_0, t) | \psi (\xi', \psi_0, t) \rangle} \right],
\end{aligned}
\end{equation}
where $\xi (\xi_0, \psi_0, t)$ and $\psi (\xi_0, \psi_0, t) \rangle$ are the classical and quantum parts, respectively, of the Ehrenfest trajectory with initial conditions $(\xi_0, \psi_0)$. The dynamical equation for the density matrix has a structure similar to the equation introduced by Truhlar and co-workers for density matrices \cite{zhu2005non, Truhlar2007Book}:
\begin{equation}
\label{dissipTruhlar}
\dot{\rho} (t) = \dot{\rho}^C (t) + \dot{\rho}^D (t),
\end{equation}
where $\dot{\rho}^C (t)$ models the coherent evolution (i.e evolution according to the Liouville-von Neumann equation) and $\dot{\rho}^D (t)$ represents the dissipation term. By comparing this decomposition with the dynamical equation \eqref{eq:22}, we can conclude that the description of hybrid systems with uncertainty evolving under ESD allows us to obtain an {\it a priori} formulation for dissipative terms. In Sections \ref{secDefSys} and \ref{secResults}, the particular characteristics of decoherence in ESD for a simple molecular system will be described.

\subsubsection{Purity change, decoherence time, and stable projectors}
\label{S2.3}

Consider an isolated quantum system. The evolution of its density matrix $\rho$ is governed by von Neumann's equation. In this case, its purity,
\begin{equation}
\label{eq:12}
{\mathcal P} (\rho) = \mathrm{Tr} (\rho^2),
\end{equation}
is a constant of motion. For a (purely quantum) composite system, however, the situation is different. The evolution of the reduced density matrix of a subsystem, obtained through a partial trace on the total one, does not verify in general von Neumann's equation, and its purity is no longer conserved. An initial pure state, therefore, evolves into an statistical mixture of pure states, i.e. a mixed state. In some circumstances, an equation of motion can be derived or assumed for the reduced density matrix (e.g. of Lindblad type), describing the coupling of the reduced system to an external bath that may cause not only changes in purity, but also decoherence.

The statistical hybrid model described above is also a composite system, although one of the subsystems is no longer quantum but classical. We should expect to find changes in purity and presence of decoherence, as in the case of purely quantum composite systems. In the context of environment-induced decoherence \cite{Paz1999a}, decoherence in an open quantum system occurs when the interaction with an environment causes a non-reversible evolution of the quantum state of the system, represented by a density matrix, whose non-diagonal terms vanish for long times. Physically, the system typically evolves from a single state to a statistical mixture of states, whose projectors form the spectral decomposition of the final density matrix. One of the key points in the description of decoherence is that these states are not arbitrary: the \emph{decoherence hypothesis}\cite{Dannenberg2011, Lychkovskiy2009, Schlosshauer2007} states that these are elements from a fixed set selected by the environment, called the \emph{pointer basis}. Thus, the presence of decoherence in the evolution of a quantum system can be deduced from the identification and description of a pointer basis for the system.

More specifically, in the context of HQCD, decoherence may appear by analising the evolution of the quantum subsystem; in this case, the classical part of the system acts as the environment that may select the pointer basis for the quantum subsystem. Due to the particular mathematical description of these models, decoherence in HQCD is due to the fact that neglected wave functions in the classical limit rapidly lose overlap in time. The pointer basis of the system is formed because the evolution of the classical subsystem destroys the superposition of states in the quantum subsystem\cite{Bedard2005JCP, Larsen2006, Dannenberg2011, Lychkovskiy2009, Schlosshauer2007}. In conclusion, the interaction described in Ehrenfest dynamics causes the existence of a pointer basis for the quantum subsystem, selected by the classical subsystem, consisting of projectors onto subspaces of the Hilbert space of the quantum system which are both stable and attractors of the dynamics of the density matrices.

It is noticeable that our approach, which does not depend on any particular choice of basis for the system, allows us to consider a definition of \emph{decoherence time} $\Delta$ which is completely intrinsic. As the decrease in purity is a characteristic of decoherent evolution, we will consider that, for a system loosing coherence it time, such decoherence time is reached when its purity ${\mathcal P}$ stabilises.

In order to determine the pointer basis of a system, two aspects will be considered: stability and attractiveness. Consider a hybrid system evolving from a pure state and  with some uncertainty on the initial values of the classical degrees of freedom. If decoherence occurs, then after a period determined by the decoherence time, the system will have evolved into an approximately stable mixture of states. Thus, in each particular example, the pointer basis can be identified as the set of those stable states, if any, whose projectors conform the density matrix of the system. In the next section, a simple diatomic molecule will be analysed both analytically and numerically, and we will determine the pointer basis of this molecular system.

\section{Example: dynamical simulation of an ionised dimer with uncertainty}
\label{sec:exampl-dynam-simul}

\subsection{Definition of the system}
\label{secDefSys}

We have chosen as an example an ionised dimer, a simple yet non trivial model of molecule. For simplicity, we consider two atoms with only one valence electron, and all the inner electrons are identified together with the corresponding nucleus as a single classical particle called a `core'. Summing up, the whole ionised molecule is described by two classical cores of charge +1 and a single quantum valence electron. Each core is given a mass of 23 a.m.u.; in this way, the system could be seen as a simple model for a ionised dimer of sodium, Na$_2^+$.

We used the Octopus software \cite{MARQUES200360, Andrade2015}, that uses an intuitively simple grid-based basis to represent the quantum degrees of freedom. The eigenstates and eigenvalues of the electronic Hamiltonian $H_e(R)$, as well as the dynamics, are computed in regular rectangular 3-dimensional grid contained in a spherical simulation box. The convergence tests showed that a grid with radius 13 a.u. and spacing 1.2 a.u is dense enough for a precise description of the classical and quantum degrees of freedom of our molecule. The dimension of the Hilbert space is thus given by the number of points in the real space grid. We did not impose absorbing boundary conditions at the walls of the chamber, since the simulations did not lead the electronic cloud to approach them.

\begin{figure*}[t]
\centering
\begin{subfigure}{0.32\textwidth}
\includegraphics[width=\textwidth]{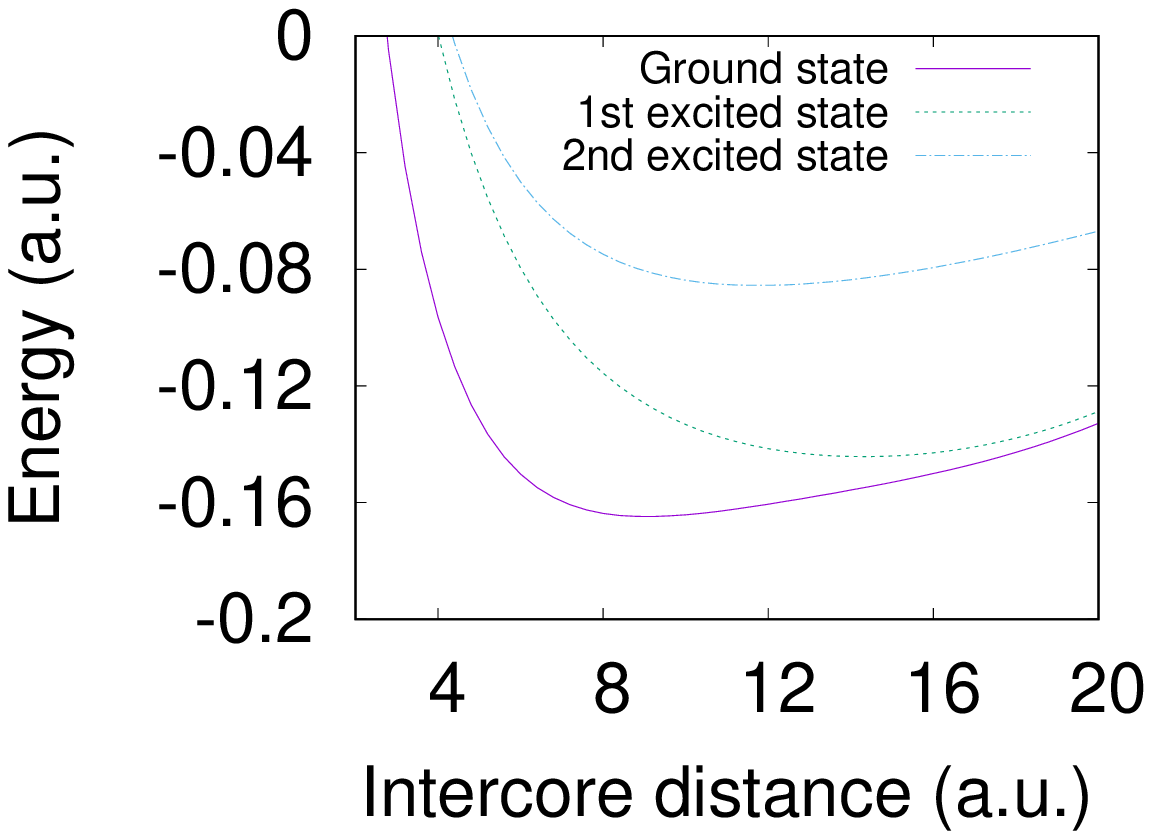}
\caption{}
\label{FigEigenHeR}
\end{subfigure}
\begin{subfigure}{0.32\textwidth}
\centering
\includegraphics[width=\textwidth]{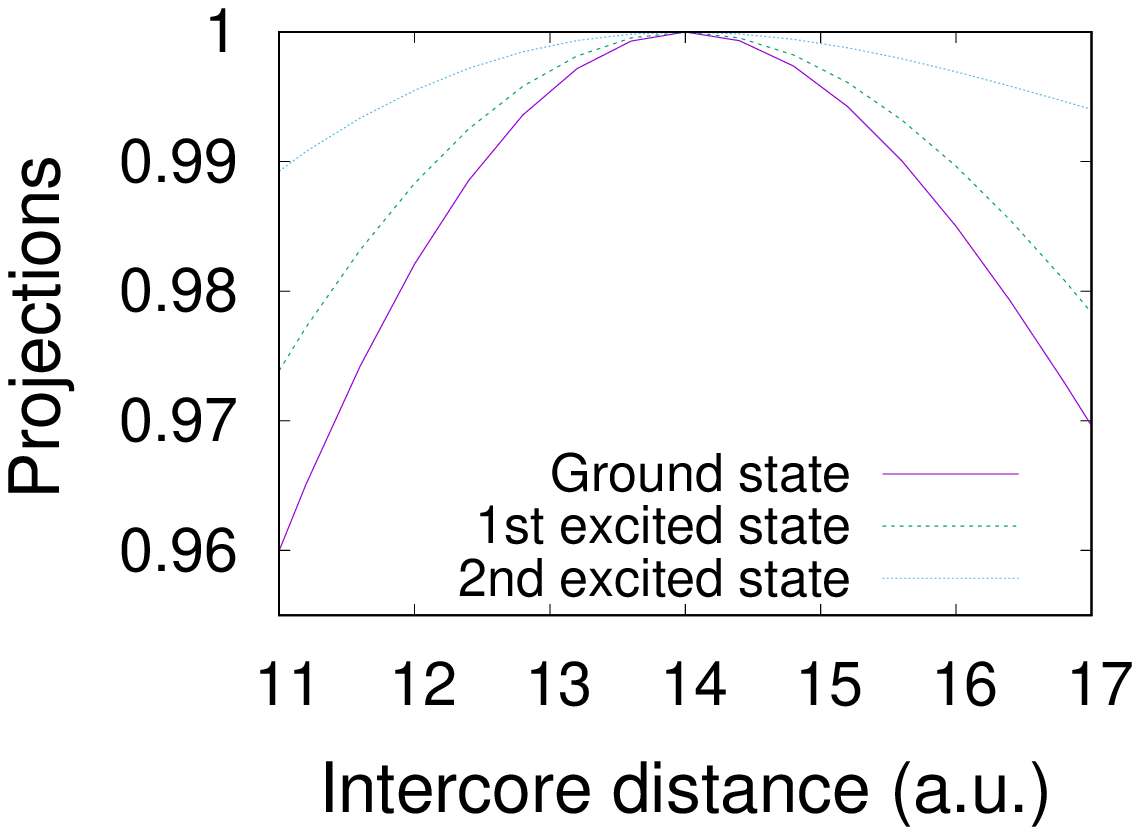}
\caption{}
\label{figProjEigenV}
\end{subfigure}
\begin{subfigure}{0.32\textwidth}
\includegraphics[width=\textwidth]{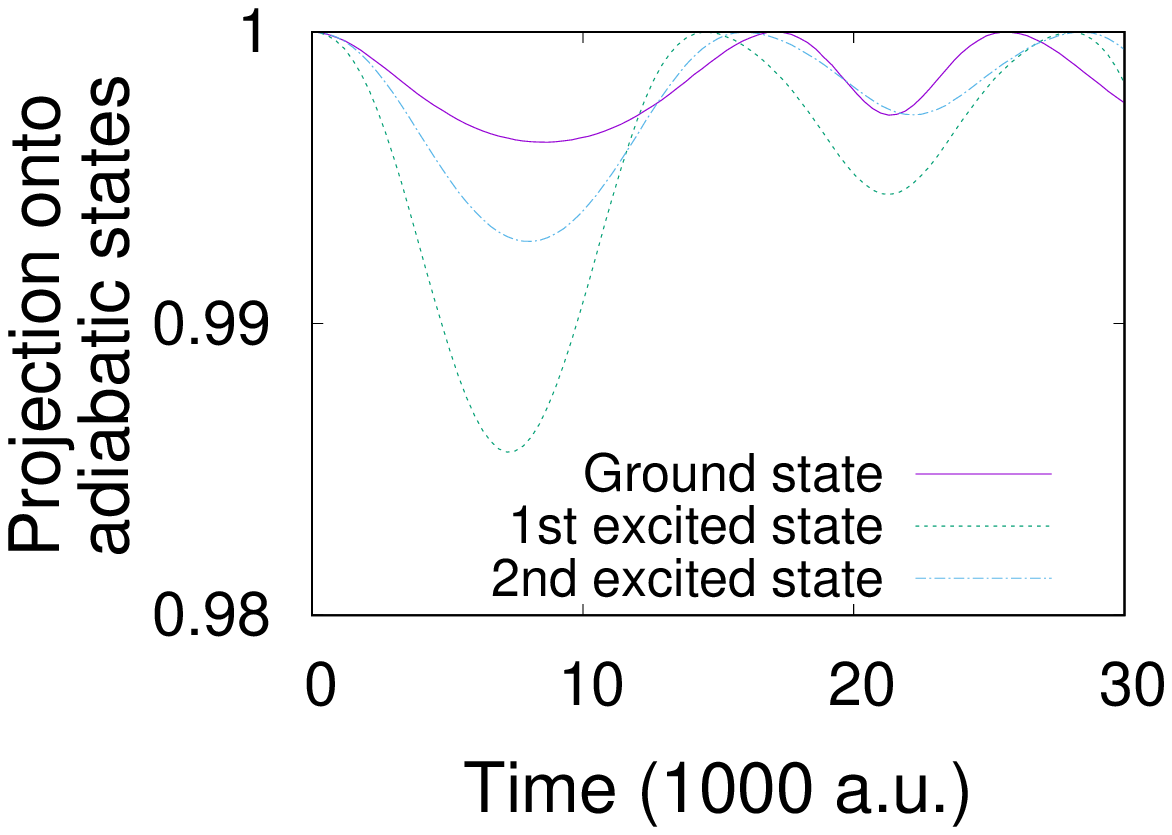}
\caption{}
\label{figAdiabatic}
\end{subfigure}
\caption{{\bf (a)} Eigenvalues of the electronic Hamiltonian. {\bf (b)} Projection $|\langle \phi_j (R)|\phi_j(R_0) \rangle|$ of eigenstates $\phi_j(R)$ of the electronic Hamiltonian $H_e(R)$ onto the eigenstates of $H_e(R_0)$, with $R_0$ a position for the cores along the $x$ axis and separated 14 a.u. All the computations are restricted to positions of the cores along the $x$ axis. {\bf (c)} Stability of eigenstates of the electronic Hamiltonian under the evolution given by the Ehrenfest model. For initial quantum states $\psi^{(j)}(0)=\phi_j(R_0)$, the figure represents the projections $|\langle \phi_j(R(t)) |\psi^{(j)} (t)\rangle|$, for $j=0,1,2$. It can be concluded that the system follows approximately an adiabatic evolution.
}
\end{figure*}

The ion-electron interaction in Octopus is handled with pseudo-potentials, which avoid the divergence of the real Coulomb interaction. In this case, we opted for a simple soft-Coulomb potential, whose expression for a particle with charge $Z$:
\begin{equation}
\label{pseudoV}
V_\alpha(r) = \frac{Z}{\sqrt{\alpha^2 + r^2}}, \quad \alpha > 0,
\end{equation}
We take $Z=1$ for the simulation of the ionised dimers. We have chosen the value $\alpha = 3$ a.u., which reproduces approximately the experimental properties of the Na$_2$ neutral molecule. With this new potential, it is possible to consider the eigenvalue equation for the electronic Hamiltonian:
\begin{equation}
\label{HeRspecDecomp}
H_e(R) |\phi_j(R) \rangle = E_j(R) |\phi_j(R) \rangle, \ j = 0,1,2\ldots
\end{equation}
with $E_0(R) \leq E_1(R) \leq \ldots$ The eigenstates of the electronic Hamiltonian are assumed to be normalised. Figure \ref{FigEigenHeR} shows the first eigenvalues of this operator. For each set $R$, the eigenvectors of the electronic Hamiltonian form a suitable basis for the Hilbert space of electronic states.

It is possible to simplify the problem by restricting the positions of the cores. In the following, the set of positions of the cores, hereafter denoted by $\mathcal{D}$, is restricted to positions $R$ along the $x$ axis, and such that the intercore distance varies only between 11 and 17 a.u. Figure \ref{figProjEigenV} shows the behaviour of the eigenstates of $H_e(R)$: by taking a reference position $R_0$, the change in the $j$-th eigenstate is given by the projection $| \langle \phi_j(R) | \phi_j(R_0) \rangle|$. As the figure suggests, it can be proved that
\begin{equation}
\label{projR}
\langle \phi_j(R) | \phi_k (R') \rangle \simeq \delta_{jk}, \quad \forall R, R' \in \mathcal{D}.
\end{equation}

It is useful for the problem at hand to consider a basis for the electronic states formed by eigenstates of the electronic Hamiltonian at a reference position $R_0$ of the cores:
\begin{equation}
\label{basisB}
\mathcal{B} = \{\phi_j := \phi_j (R_0) \mid j=0,1 \ldots \}.
\end{equation}
For any $R \in \mathcal{D}$, the electronic Hamiltonian $H_e(R)$ is approximately diagonal in this basis, which simplifies the computations.

The next step in the description of the problem is the analysis of the dynamics determined by the Ehrenfest model \eqref{eq:ehrenfest}. For given initial conditions $(R_0, P_0, \psi_0)$, integration of the Ehrenfest model gives the evolution $(R(t), P(t), \psi(t))$ of the quantum and classical degrees of freedom. Initial parameters $R_0$ and $P_0$ are taken along the $x$ axis, so that $R (t) \in \mathcal{D}$ for all $t\geq 0$.

With these tools, it is interesting to analysis the adiabaticity of the Ehrenfest model when applied to the described example. For this purpose, let us choose as initial quantum state any of the eigenstates of the electronic Hamiltonian at $R_0$, i.e. $\psi^{(j)} (0) = \phi_j (R_0)$. We have computed numerically the projections $\langle \phi_k (R(t)) |\psi^{(j)}(t)\rangle$ for different values of $t$ Figure \ref{figAdiabatic} represents these projections for $j=k=0,1,2$. We have found that these projections take approximately value 1 when $j=k$. As for each $R \in \mathcal{D}$ elements in the basis $\{\phi_j(R)\}$ are orthogonal to each other, the rest of projections are approximately zero. This leads to the following approximation
\begin{equation}
\label{projAdiab}
\begin{aligned}
& \psi^{(j)}(0) = \phi_j (R_0) 
\Rightarrow \langle \phi_k (R(t)) |\psi^{(j)}(t)\rangle \simeq \delta_{jk}, \ \forall t.
\end{aligned}
\end{equation}
We can conclude that, for the considered example, the system behaves in an approximately adiabatic way.

Summing up, the two relevant approximations found for this model are \eqref{projR} and \eqref{projAdiab}. Both can be combined to approximate the evolution of generic initial conditions by ED \eqref{eq:ehrenfest} (as long as positions $R(t)$ of the cores stay in the allowed set $\mathcal{D})$:
\begin{equation}
\label{evolAdiab}
\begin{aligned}
& \psi (0) = \sum_j c_j \phi_j, \quad c_0, c_1, \ldots \in \mathbb{C}
\Rightarrow
\psi (t) \simeq \sum_j c_j \exp \left( -i \int_0^t E_j(R(t')) dt' \right) \phi_j.
\end{aligned}
\end{equation}

It is important to notice that adiabaticity (neither exact nor approximate) is not a general property of ED, as the Ehrenfest equations introduces non-adiabatic couplings \cite{Alonso2011}. If the state is written at each $t$ as
\begin{equation}
\psi (t) = \sum_j c_j (t) \phi_j (R(t)),
\end{equation}
then from the Ehrenfest equation \eqref{eq:ehrenfest} it is immediate to compute that
\begin{equation}
\begin{aligned}
i\frac{d}{dt} c_j(t) & = E_j(R(t)) c_j(t)
- i \sum_k c_k(t) \left[ \sum_J \frac{\vec{P}_J}{M_J} \cdot \vec{d}_J^{\,jk} (R(t)) \right], \\
\vec{d}_J^{\,jk} (R) & = \left\langle \phi_j(R) \left| \frac{\partial}{\partial \vec{R}_J} \phi_k(R) \right. \right\rangle.
\end{aligned}
\end{equation}
with $\vec{d}_J^{\,jk} (R)$ the non-adiabatic couplings. However, in our examples, the second term in the right hand side turns out to be negligible.

\subsection{Statistical uncertainty for a single dimer: Decay of purity and pointer basis}
\label{secCuentas}

We are interested in the appearance of decoherence effects: purity changes and pointer basis. For this reason, let us consider that the uncertainty in the initial conditions of our dimer is described by the following probability density function:
\begin{equation}
\label{eq:6}
\begin{aligned}
F_{QC}(\xi, \psi;0) = & \frac 1N \left( \sum_{j=1}^N \delta(P-P_{j,0})\right)
\delta(R-R_0) \delta(\psi-\psi_0).
\end{aligned}
\end{equation}
with $R=(\bf{R}_1, \bf{R}_2)$ and $P=(\bf{P}_1, \bf{P}_2)$ the position and momenta, respectively, of the two cores in our model, and $\psi$ the quantum state of the valence electron. Observe that the only uncertainty is assumed in the classical degrees of freedom, while the quantum subsystem is taken in the initial state $\psi_0 \in M_Q$. For simplicity, the initial ionic positions are also fixed, with $R_0$ the equilibrium position of the cores along the $x$ axis for the given quantum state $M_Q$. As explained above, initial momenta $P_{j,0}$ should be taken along the $x$ axis. Numerical results show that a value of $N=41$ is high enough to provide good statistical results.

The evolution of the probability distribution can be written as
\begin{equation}
\label{eq:7}
\begin{aligned}
F_{QC} (\xi, \psi; t) = & \frac 1N \sum_{j=1}^N \Big( \delta (R-R_j(t)) \delta (P-P_j(t)) \ \delta (\psi-\psi_j(t)) \Big),
\end{aligned}
\end{equation}
From this expression we can write the corresponding density matrix as:
\begin{equation}
\label{eq:5b}
\begin{aligned}
\rho(t) = & \int_{M_C \times M_Q} d\mu_{QC} (\xi, \psi) F_{QC} (\xi, \psi;t) \frac{|\psi\rangle \langle \psi|}{\langle \psi|\psi \rangle} 
= \frac 1N \sum_{j=1}^N \frac{|\psi_j (t)\rangle \langle \psi_j (t)|}{\langle \psi_j (t)|\psi_j (t)\rangle}.
\end{aligned}
\end{equation}
Let us now analyse some simple examples, which will become useful in order to predict the changes in purity and the existence of a pointer basis for the molecule.

As a first approach to the problem, consider in the initial probability distribution \eqref{eq:6} that the initial quantum state is an eigenstate of the electronic Hamiltonian, i.e. $\psi_0=\phi_k \in \mathcal{B}$. An approximation to the density matrix of the system can be obtained by substituting \eqref{evolAdiab} in \eqref{eq:5b} for every set of initial conditions $(R_0, P_{j,0}, \psi_0)$. Observe that, due to the small variation of the eigenstates of $H_e(R)$ and to the approximate adiabaticity of the dynamics, eigenstates of the electronic Hamiltonian are approximately constant (except for a global phase). As a consequence, the purity of the quantum subsystem is close to 1 for all the evolution. Observe that, although no decoherence seems to occur in this example, this is the first hint of the existence of a pointer basis for the molecule.

\begin{figure*}
\centering
\begin{subfigure}{0.32\textwidth}
\includegraphics[width=\textwidth]{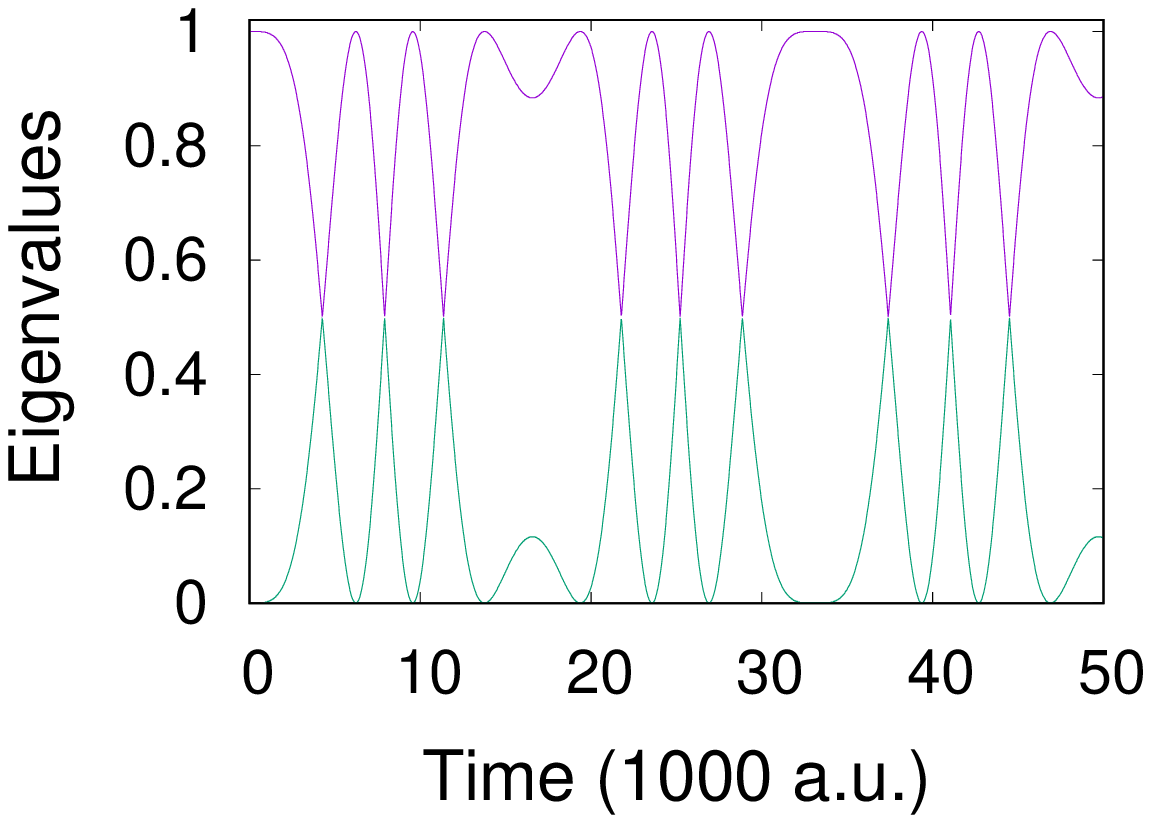}
\caption{}
\label{fig:2traj-eigenvaluesLinear}
\end{subfigure}
\begin{subfigure}{0.32\textwidth}
\includegraphics[width=\textwidth]{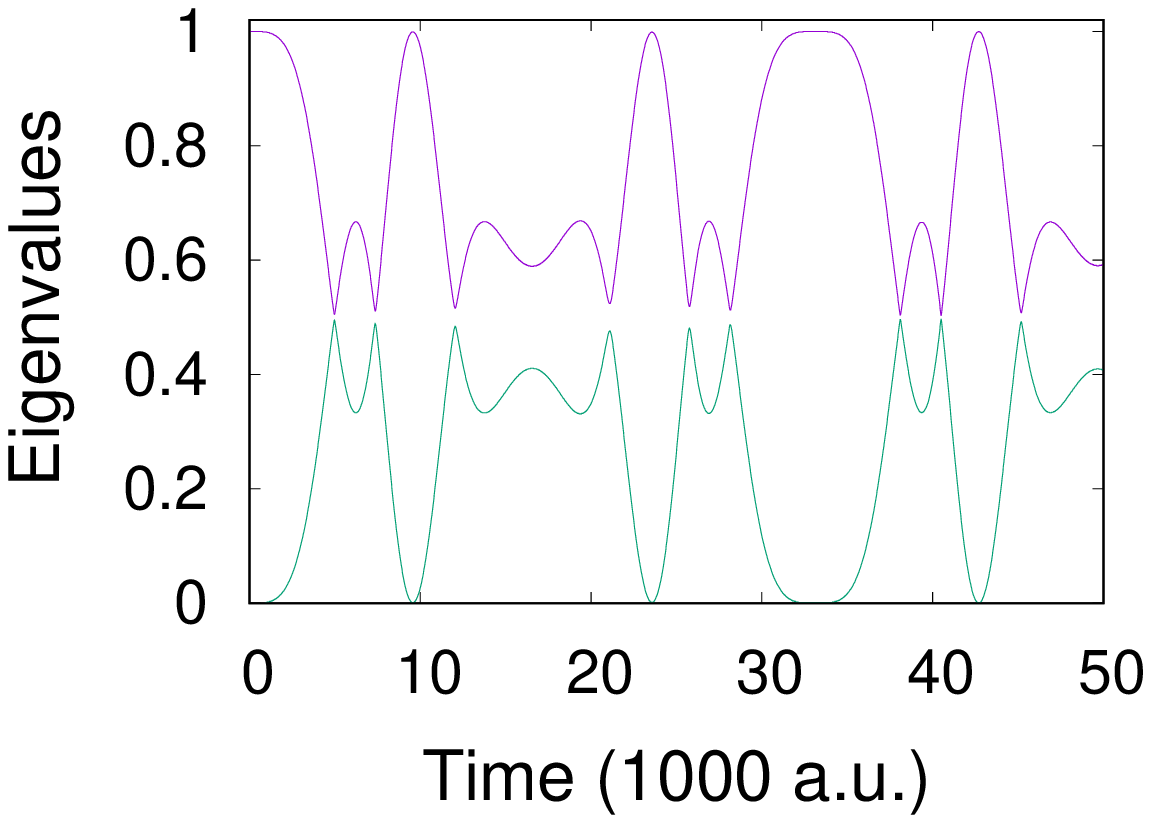}
\caption{}
\label{fig:3traj-eigenvaluesLinear}
\end{subfigure}
\begin{subfigure}{0.32\textwidth}
\centering
\includegraphics[width=\textwidth]{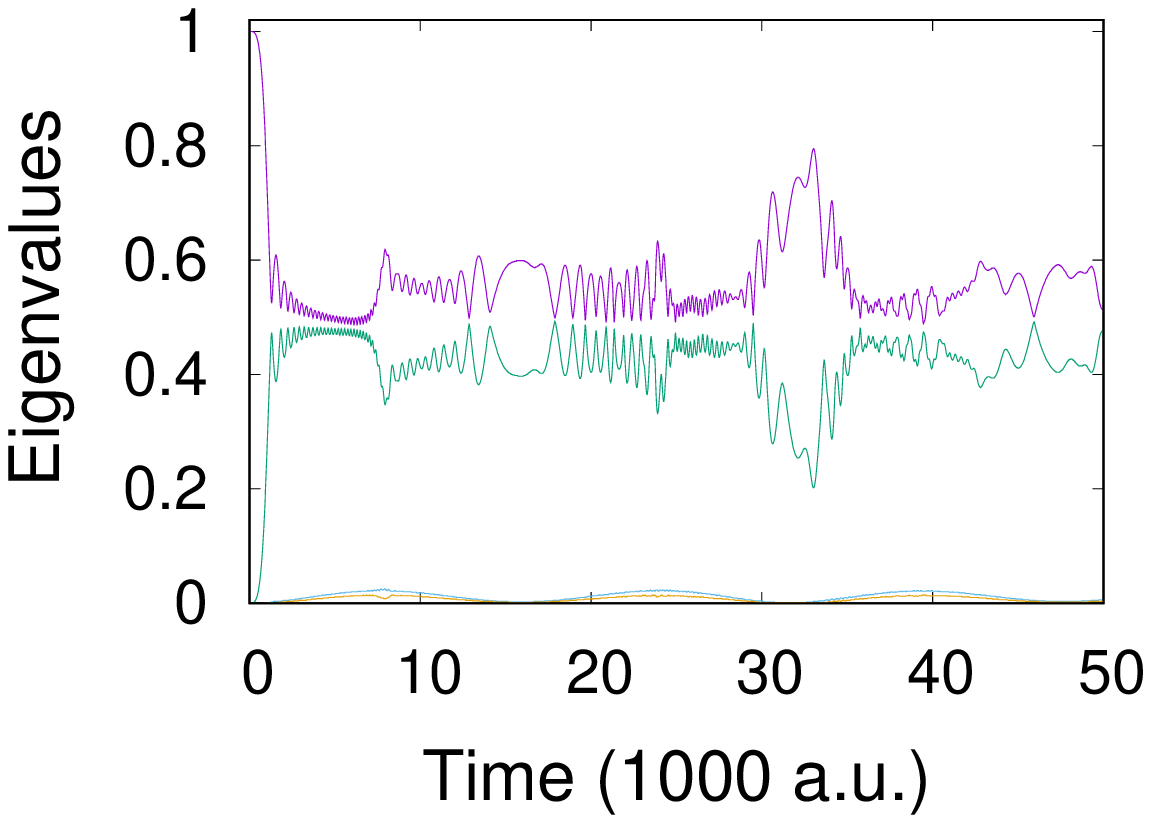}
\caption{}
\label{fig:2-eigenvalues}
\end{subfigure}
\caption{{\bf (a)} Evolution of eigenvalues of $\rho(t)$ for the case of a quantum initial condition corresponding to 2 eigenstates and $N=2$. {\bf (b)} Evolution of eigenvalues of $\rho(t)$ for the case of a quantum initial condition corresponding to 2 eigenstates and $N=3$. {\bf (c)} Evolution of the eigenvalues of $\rho(t)$ for the case of a quantum initial condition corresponding to 2 eigenstates and $N=41$. Notice that small non-zero eigenvalues appear also in this case.
}
\end{figure*}

Consider now a second example. In order to extract some conclusion on the behaviour of the system, we choose $N=2$ and we take the following expression for the initial probability distribution:
\begin{equation}
\label{iniEx2}
\begin{aligned}
F_{QC} (R,P,\psi; 0) = & \frac 12 \Big( \delta (P-P_{1,0}) + \delta (P-P_{2,0}) \Big)
\delta (R-R_0) \, \delta \left( \psi - \psi_0 \right),
\end{aligned}
\end{equation}
where the initial quantum state is chosen as
\begin{equation}
\psi_0 = \frac{1}{\sqrt{2}} \left( \phi_0 + \phi_1\right).
\end{equation}
According to \eqref{evolAdiab}, the quantum trajectories $\psi_1(t), \psi_2(t)$, determined by the Ehrenfest equations with initial conditions $(R_0, P_{1,0}, \psi_0), (R_0, P_{2,0}, \psi_0)$ respectively, are the following
\begin{equation}
\label{eq:40}
\begin{aligned}
\psi_j (t) \simeq & \frac{1}{\sqrt{2}} \exp \left( -i \int_0^t E_0(R_j(t')) dt' \right) \phi_0
+ \frac{1}{\sqrt{2}} \exp \left( -i \int_0^t E_1(R_j(t')) dt' \right) \phi_1,
\end{aligned}
\end{equation}
for $j=1,2$. Substituting in \eqref{eq:5b} it is possible to estimate the density matrix of the quantum subsystem:
\begin{equation}
\begin{aligned}
\rho(t) \simeq & \frac 12 \left( |\phi_0\rangle \langle \phi_0| + |\phi_1\rangle \langle \phi_1|
+ \frac{e^{i \Delta_1 (t)} + e^{i \Delta_2 (t)}}{2} |\phi_0\rangle \langle \phi_1|
+ \frac{e^{-i \Delta_1 (t)} + e^{-i \Delta_2 (t)}}{2} |\phi_1\rangle \langle \phi_0| \right),
\end{aligned}
\end{equation}
where we have defined the quantities
\begin{equation}
\Delta_j (t) = \int_0^t g_j (t') d t', \quad
g_j (t) = E_1(R_j(t)) - E_0(R_j(t)),
\quad j=1,2.
\end{equation}
Observe that $g_j (t)$ are the gaps between the ground state and the first excited state energies for each initial conditions at each time $t$. Notice that the density matrix $\rho(t)$ corresponds to the sum of two projectors, each one projecting onto the subspace generated by the vector
\begin{equation}
\label{eq:44}
\chi_j (t) = e^{-i \Delta_j (t)} \phi_0 + \phi_1, \quad j=1,2.
\end{equation}
Even if the difference $|g_1(t) - g_2(t)|$ between the gaps is very small, integration in time makes the exponent $\Delta_j(t)$ non-negligible. Thus, the difference between the vectors $|\psi_1 (t)\rangle$ and $|\psi_2 (t)\rangle$ will become periodically relevant for the description of the quantum state. We can verify this assumption by computing the spectral decomposition of the density matrix $\rho(t)$, defined in \eqref{eq:5}. The dependence of its eigenvalues with time is presented in Figure \ref{fig:2traj-eigenvaluesLinear}. We notice that the number of eigenvalues different from zero is indeed a function of time, depending on the orthogonality of vectors $\chi_1 (t)$ and $\chi_2 (t)$.

A more general analysis can be done by extending the above computation to generic values of $N$ in the initial distribution \eqref{iniEx2}. If the quantum state is approximated by \eqref{evolAdiab}, the density matrix of the quantum subsystem has the following form:
\begin{equation}
\label{rho40Sum}
\begin{aligned}
& \rho (t) \simeq \frac 12 \left( |\phi_0\rangle \langle \phi_0| + |\phi_1\rangle \langle \phi_1| \right) + \frac{1}{2N} \sum_{j=1}^{N} \left( e^{-i \Delta_j (t)} |\phi_0\rangle \langle \phi_1| + e^{i \Delta_j (t)} |\phi_1\rangle \langle \phi_0| \right),
\end{aligned}
\end{equation}
which is a sum of projectors onto the subspaces generated by
\begin{equation}
\label{eq:44b}
\chi_j(t) = e^{-i \Delta_j} \phi_0 + \phi_1, \quad j=1, \ldots , N.
\end{equation}

Notice that the corresponding sum of projectors tend to zero when $N$ grows, since we are adding $N$ time-dependent vectors of norm one, moving with different velocities, in the linear space generated by $\phi_0$ and $\phi_1$. The case for $N=3$ is represented in Figure \ref{fig:3traj-eigenvaluesLinear}. The case $N=41$ is represented in Figure \ref{fig:2-eigenvalues}. Notice how we obtain two eigenvalues approximately equal and a series of remaining eigenvalues representing the effect of the change of the position of the cores in the spectrum of the Hamiltonian $H_{e}(R)$ and the non-vanishing part from the projector sum, with a negligible weight. Indeed, for large values of $N$, coefficients in the sum in \eqref{rho40Sum} are an approximately random set of complex numbers with modulus one, their sum being zero. Thus, for large enough values of $t$, the density matrix tends to
\begin{equation}
\label{eq:46}
\rho(t) \to \frac 12 \left( |\phi_0\rangle \langle \phi_0| + |\phi_1\rangle \langle \phi_1|\right),
\end{equation}

The asymptotic value of the purity for a generic case can be computed based on the identification of the pointer basis with eigenstates of the electronic Hamiltonian, as detailed above. If the initial state is a linear superposition of $n$ such states, with equal coefficients, then the purity tends to a value of $1/n$. For generic linear combinations, the asymptotic value of purity can be computed as:
\begin{equation}
\label{purLim}
\begin{aligned}
\psi_0 = & \sum_j c_j \phi_j, \ c_0, c_1, \ldots \in \mathbb{C} 
\Rightarrow \rho(t) \to \sum_j |c_j|^2 |\phi_j \rangle \langle \phi_j|, \quad
\mathcal{P} (\rho(t)) \to \sum_j |c_j|^4.
\end{aligned}
\end{equation}
Observe that the complex phase of the coefficients is irrelevant for the computations, as the eigenstates of the electronic Hamiltonian can always be redefined as $c_j \phi_j = |c_j| \phi'_j$.

Additionally, the above computations show the existence of pointer basis in ESD. According to the meaning of decoherence in the context of HQCD \cite{Bedard2005JCP, Larsen2006}, described above, the pointer basis appears as the set of stable states under the non-unitary evolution of the system. From \eqref{evolAdiab}, it can be concluded that eigenstates of the electronic Hamiltonian, i.e. elements in the basis $\mathcal{B}$, are for ESD the elements in the pointer basis of the quantum system.

Summarising, we have observed that the classical subsystem (the cores), acting as an environment onto the valence electron, produce a dynamics that, for times longer than the decoherence time, selects a small set of the possible quantum states. In this particular example, these are the eigenstates of the electronic Hamiltonian. These conclusions can be a starting point for a full analysis of pointer basis in molecular systems and in ESD, which could be performed by considering more general initial parameters and larger molecular systems. Observe that this is in agreement with the decoherence hypothesis \cite{Lychkovskiy2009}, as the pointer basis does not depend on the initial state of the quantum subsystem. It is also important to notice that the results here obtained are characteristic of ESD. Other decoherent mechanisms will present different features and the pointer basis might appear in a completely different fashion.

\subsection{Numerical simulations of an ionised dimer}
\label{secResults}

In order to illustrate the results in Section \ref{secCuentas}, we have computed explicitly the changes in the purity of a simple model of an ionised dimer, as described above. As initial conditions, we take a certain initial state $\psi_0$ of the quantum subsystem. For this state, the equilibrium position $R_0$ of the cores are taken as their initial positions. The nuclei are allowed to move only along the axis of the molecule and keeping their center of mass stationary. Some uncertainty is allowed in the values of the initial momenta, ranging from 0 to $40\times 10^{-5}$ a.u., in both directions. With these values, the kinetic energy of the cores is never large enough to dissociate the dimer. Thus, different regimes of oscillation are considered. In conclusion, the uncertainty in the initial conditions is given in the form of \eqref{eq:6}.

Three different choices for the quantum subsystems are considered. In each case, the initial position of the cores is always the equilibrium position for the given quantum state:
\begin{itemize}
\item Case A: The initial quantum state is an eigenstate of the electronic Hamiltonian:
\begin{equation}
\label{psi0A}
\psi_0 = \phi_1.
\end{equation}
\item Case B: The initial state is a linear superposition of two eigenstates of the electronic Hamiltonian:
\begin{equation}
\label{psi0B}
\psi_0 = \frac 1{\sqrt{2}} \left (\phi_0  + \phi_1 \right),
\end{equation}
\item Case C: The initial state is a linear superposition of three eigenstates of the electronic Hamiltonian:
\begin{equation}
\label{psi0C}
\psi_0 = \frac 1{\sqrt{3}} \left (\phi_0 + \phi_1 + \phi_2 \right),
\end{equation}
\end{itemize}

\begin{figure*}[t]
\begin{tabular}{cc}
	\begin{subfigure}{0.6\textwidth}
	\includegraphics[width=\textwidth]{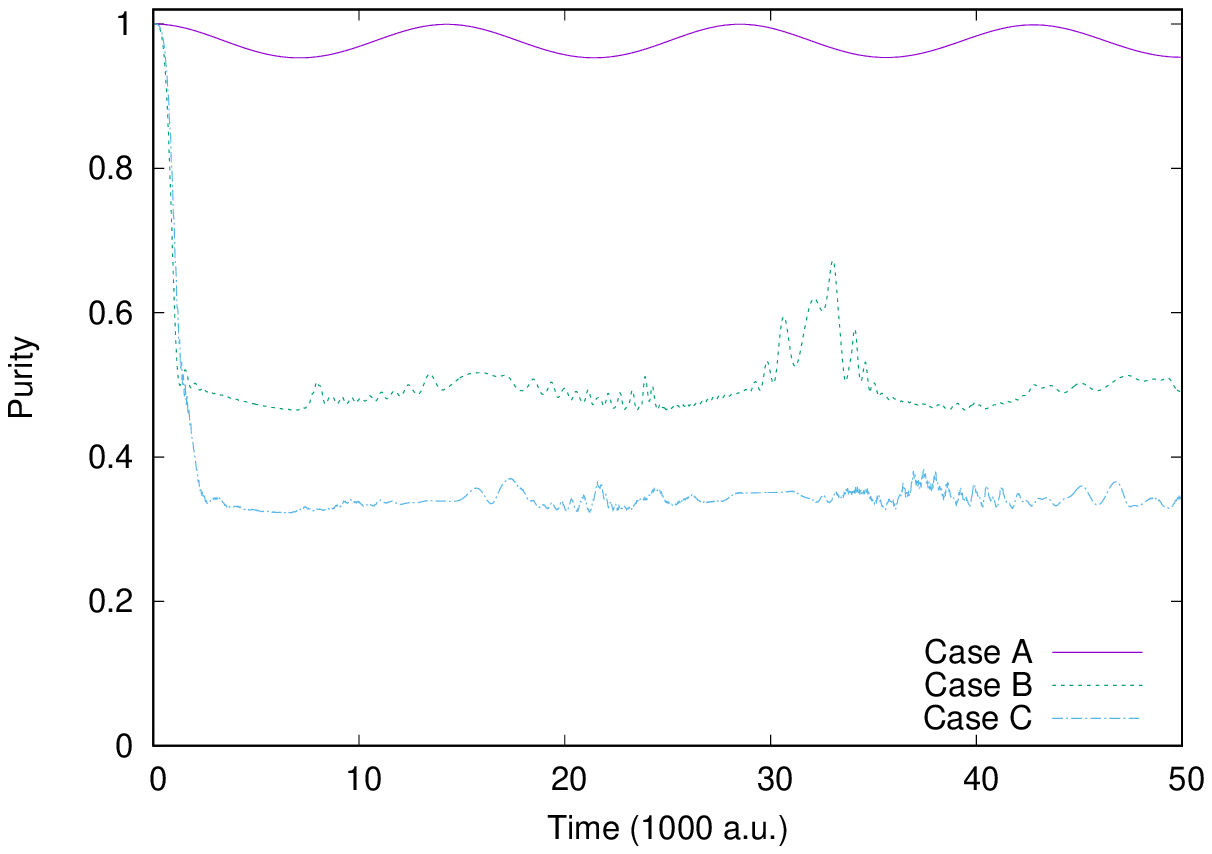}
	\caption{}
	\label{figPurities}
	\end{subfigure}
	&
	\begin{tabular}{c}
		\begin{subfigure}{0.3\textwidth}
		\includegraphics[width=\textwidth]{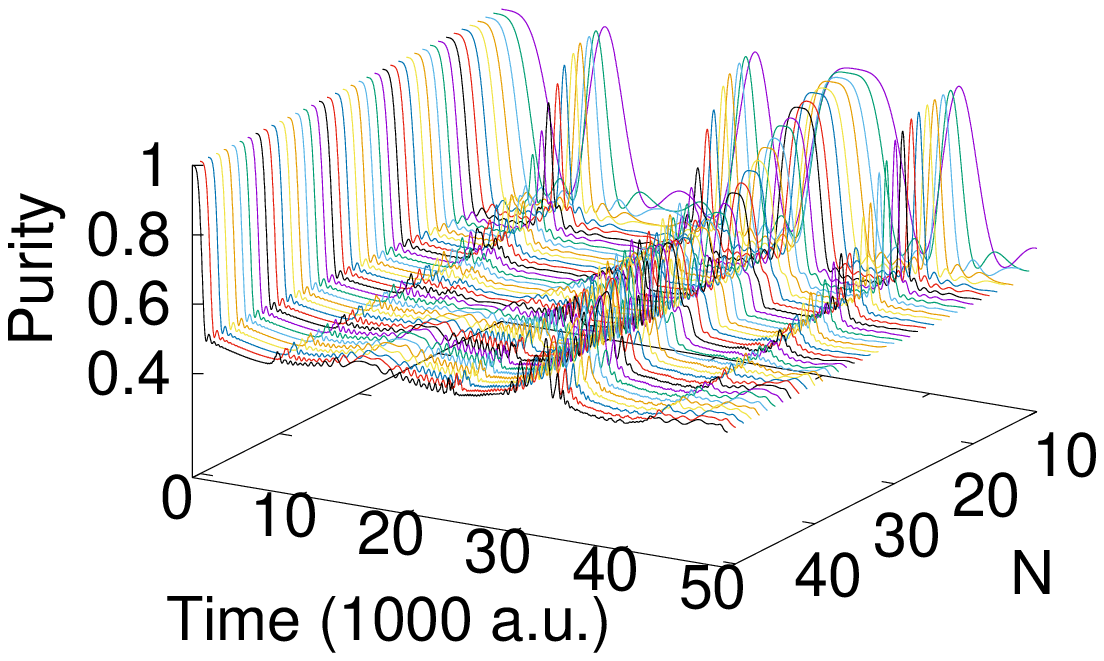}
		\caption{}
		\label{figPurProg02}
		\end{subfigure}
		\\
		\begin{subfigure}{0.3\textwidth}
		\includegraphics[width=\textwidth]{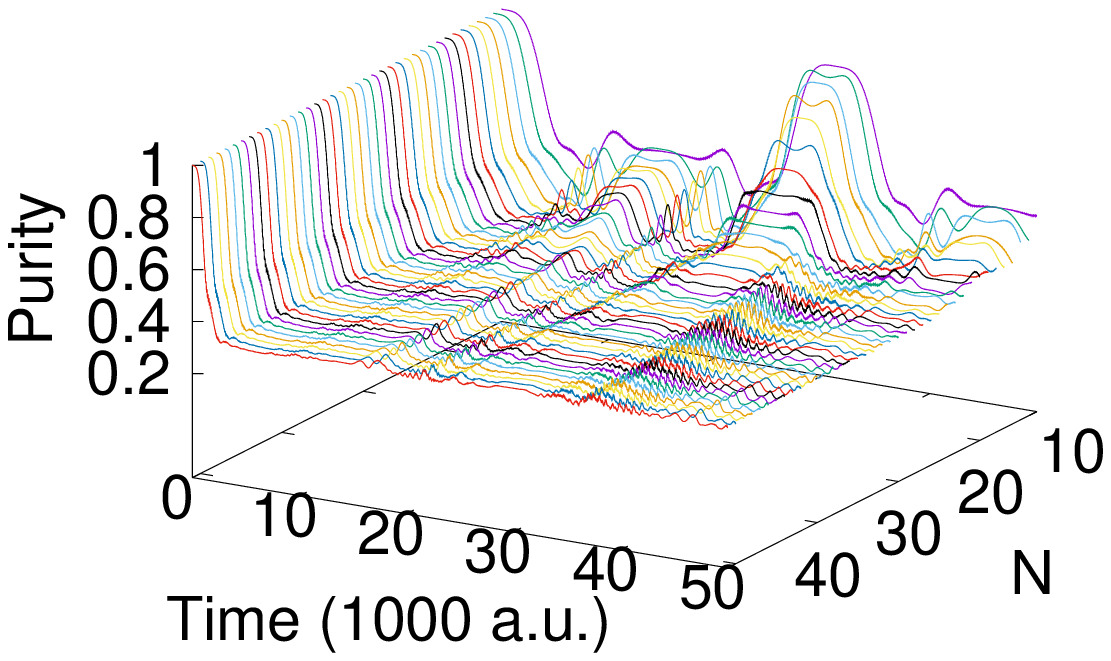}
		\caption{}
		\label{figPurProg03}
		\end{subfigure}
	\end{tabular}
\end{tabular}
\caption{Three different cases, labelled A, B and C, are considered. For each case, the initial state of the valence electron is fixed by \eqref{psi0A}, \eqref{psi0B} and \eqref{psi0C}, respectively. The cores are initially at the equilibrium position for the corresponding electronic state, and with different initial speeds along the molecule axis, as described in the text. The system evolves according to ESD. {\bf (a)} Evolution of the purity of the density matrix $\rho(t)$ for the proposed cases. The purity reaches an approximately stable asymptotic value of $1/n$, with $n$ the number of eigenstates of the electronic Hamiltonian whose linear superposition determine the initial state of the quantum subsystem. {\bf (b-c)} Evolution of the purity of the density matrix $\rho(t)$ with respect to time and the value of $N$ in \eqref{eq:6}. The initial quantum state corresponds to \eqref{psi0B} in {\bf (b)} and to \eqref{psi0C} in {\bf (c)}. It can be observed that the decoherence time and the asymptotic value of the purity stabilise for larger values of $N$.}
\end{figure*}

The results plotted in Figure \ref{figPurities} are in agreement with the behaviour given in \eqref{purLim}. ESD leads to changes in the purity of the quantum states \cite{Alonso2011, Alonso2012a}, unlike standard ED, which was purity-preserving. Let us analyse in detail each example. Case A reproduces the case in which the initial quantum state is an eigenstate of the electronic Hamiltonian at $R_0$ (see the beginning of Section \ref{secCuentas}). The purity is approximately constant and equal to 1 for all the evolution. On the contrary, cases B and C show a sharp drop on the purity (on the left side of Figure~\ref{figPurities}), and reach approximately stationary values of 1/2 and 1/3, respectively.

The decoherence time can be read from the plots as the time it takes for the system to reach a stationary value. This definition is of course not precise, but in any case this time is approximately 1200 and 2500 a.u. for cases B and C, respectively. This time is very short in comparison with the typical oscillation time of our molecule, which has been computed to be approximately 30000 a.u.

Observe that the numbers $N$ of different trajectories considered in the mixtures are essential in order to understand the behaviour of the system. Figures \ref{figPurProg02} and \ref{figPurProg03} represents the purity of cases B and C, respectively, for different numbers of trajectories, chosen in increasing order of initial speeds of the cores. It can be inferred that a low number of trajectories causes large oscillations in the purity, while adding trajectories causes the stabilisation of the value of the purity after the decoherence time. If much large cases were considered, these oscillations are expected to disappear. Decoherence time would then determine when the system has become a mixture of states from the pointer basis, and the decoherence hypothesis introduced in Section \ref{S1} is satisfied \cite{Lychkovskiy2009}.

\section{Conclusions}
\label{secConcl}

Non-adiabatic transitions and decoherence are two of the most important phenomena in chemical dynamic reactions. To which extent does a given theoretical model include each effect is a complicated question. Several methods have been developed to deal with both notions \cite{Yonehara2012}. 

In previous works \cite{Alonso2011, Alonso2012a}, we introduced a geometric route to define the Ehrenfest Statistical Dynamics (ESD). We proved that, for a simple toy-model, the resulting ESD is purity non-preserving. Now, after having carefully defined the meaning of decoherence in the context of hybrid quantum-classical dynamics (HQCD), we have tested out that, in ESD, decoherence does appear.

In fact, in this article we have proved that, when applied to a realistic molecular model, ESD defines a noncoherent evolution for the electronic degrees of freedom of the molecule. In our computations, the state space encoding the quantum degrees of freedom is considered as the finite dimensional vector space obtained by considering the values of the electronic wave-function on a grid, instead of the more usual approach for Ehrenfest dynamics that considers the truncated expansion in the electronic Hamiltonian eigenbasis. It is important to notice that this choice and the intrinsic (i.e., tensorial) nature of our formalism ensures that our results are completely independent of any choice of basis for the system. Thus, in this intrinsic way, we have proved how the statistical nature of the model is able to encode, in the simple language of Ehrenfest dynamics, non-coherent phenomena which are observed in realistic situations.
  
Some features of the decoherence phenomenon have been investigated. In particular, the purity of the electronic state has been observed to decrease in a short time, reaching an asymptotic value. Also, we have observed the existence of a pointer basis, which in our example turns out to be composed of the eigenstates of the electronic Hamiltonian of the molecular system. Notice that this fact cannot be predicted for other systems and situations (i.e. situations in which the evolution is not approximately adiabatic). In conclusion, the present paper shows that decoherence and pointer basis can be observed in simple models evolving under ESD. Further studies are required in order to extend this analysis to more complex systems.

\section*{Acknowledgement}

We are deeply grateful to the referee for all his comments that, in our opinion, have helped us to greatly improve our work. The authors have received support by the research grants E24/1 and E24/3 (DGA, Spain), MINECO MTM2015-64166-C2-1-P and FIS2017-82426-P, MICINN FIS2013-46159-C3-2-P and FIS2014-55867-P. Support from scholarships B100/13 (DGA) and FPU13/01587 (MECD) for J. A. J-G is also acknowledged. Authors acknowledge the use of ``Servicio General de Apoyo a la Investigaci\'on-SAI'', Universidad de Zaragoza.

\end{document}